\documentclass[prl,twocolumn,aps]{revtex4}
\usepackage{subfigure}
\usepackage{graphicx}
\usepackage{cancel}
\usepackage{amssymb}
\usepackage{textcomp}
\usepackage{amsmath}
\usepackage{bm}
\usepackage{times}
\usepackage{epsfig}
\usepackage{color}
\begin{document}
\title{\Large Baryon and Lepton Number as Local Gauge Symmetries}
\author{Pavel Fileviez P\'erez$^{1}$}
\author{Mark B. Wise$^{2}$}
\affiliation{\\  \\ $^{1}$University of Wisconsin-Madison, Department of Physics \\
1150 University Avenue, Madison, WI 53706, USA}
\affiliation{$^{2}$ California Institute of Technology, Pasadena, CA, 91125 USA}
\date{\today}
\begin{abstract}
We investigate a simple theory where Baryon number (B) and Lepton number (L) are local gauge symmetries.
In this theory B and L are on the same footing and the anomalies are cancelled by adding  a single new fermionic generation. 
There is an interesting realization of the seesaw mechanism for neutrino masses. Furthermore there is a natural 
suppression of flavour violation  in the quark and leptonic sectors since the gauge symmetries and particle content forbid 
tree level flavor changing neutral currents involving the quarks or charged leptons. Also one finds that the stability of a 
dark matter candidate is an automatic consequence of the gauge symmetry. Some constraints and signals at the 
Large Hadron Collider are briefly discussed.  
\end{abstract}
\maketitle
\section{I. Introduction}
In the Standard Model (SM) of particle physics baryon number (B) and lepton number (L) are accidental 
global symmetries. In addition the individual lepton numbers, $U(1)_{L_i}$ where $L_i=L_e$, $L_{\mu}$, $L_{\tau}$, 
are also automatic global symmetries of the renormalizable couplings. Since neutrinos oscillate the individual 
lepton numbers $U(1)_{L_i}$, for a given $i$, cannot be exact symmetries at the electroweak scale and 
furthermore, the neutrinos are massive.  At the non-renormalizable level in the SM one can find operators 
that violate baryon number and lepton number. For example, $QQQL/\Lambda_{B}^2$ and $LLHH/\Lambda_L$, 
where $\Lambda_B$ and $\Lambda_L$ are the scales where B and L are broken, respectively~\cite{Weinberg}.  
These operators give rise to new phenomena (that are not permitted by the renormalizable couplings) 
such as proton decay~\cite{proton} and neutrinoless double beta decay~\cite{Vogel}. 

Baryon number must be broken in order to explain the origin of the matter-antimatter asymmetry 
in the Universe. Also having Majorana neutrino masses is very appealing since one can explain the 
smallness of neutrino masses through the seesaw mechanism~\cite{TypeI}. Hence lepton number 
is also expected to be broken. In the future it may be possible to observe the violation of baryon 
number and lepton number (For future experimental proposals and current bounds, see Ref.~\cite{experiments}.).

In this paper we examine the possibility that B and L are spontaneously broken gauge symmetries 
where the scale for B and L   breaking is low (i.e., around a {\rm TeV}).   Then one does not need a ``desert region"
between the weak and GUT scales to adequately suppress the contribution  of dimension six baryon number violating  operators to proton decay 
~\cite{proton}. (Note that gauging B-L does not address this issue since the dimension six operators mentioned above are B-L invariant.)  In this paper we propose a new model where B and L are gauged and are 
spontaneously broken at a low-scale. 

We construct and discuss a simple extension of the standard model  based on the gauge symmetry, 
$SU(3) \bigotimes SU(2) \bigotimes U(1)_Y \bigotimes U(1)_B \bigotimes U(1)_L$. The 
anomalies are cancelled by adding  a single new fermionic generation of opposite chirality, 
where the new leptons have ${\rm L}=3$ and the additional quarks have ${ \rm B}=1$. In this model we find that 
there is a natural suppression of flavor violation at tree level in the quark and charged leptonic sectors. 
The realization of the seesaw mechanism for the generation of neutrino masses is investigated  and we 
show that the spontaneous breaking of the gauge symmetry does not generate dangerous baryon number 
violating operators. The model has a dark matter candidate and its stability is an automatic consequence 
of the gauge symmetry. The most generic signals at the Large Hadron Collider and the relevant 
constraints are briefly discussed. 

\section{II. Gauging Baryon and Lepton Numbers}
In this section we show how to cancel the anomalies in a simple extension 
of the standard model  theory based on the gauge group 
$G_{SM} \bigotimes U(1)_B \bigotimes U(1)_L$. 
In this model the SM fields transform as:
$Q_L^T = \left(  u, \ d \right)_L \sim (3,2,1/6,1/3,0), ~
l_L^T = \left( \nu, \ e \right)_L \sim (1,2,-1/2,0,1),~ 
u_R \sim (3,1,2/3,1/3,0),~  d_R \sim (3,1,-1/3,1/3,0),$ 
and $e_R \sim (1,1,-1,0,1)$ under the gauge group. We add three generations of right handed neutrinos, 
$\nu_R\sim (1,1,0,0,1)$, to the standard model particles above to study the generation of neutrino masses. 
In order to find an anomaly free theory one needs to add additional new fermions to cancel the following anomalies 
(Of course one has to keep in mind that the anomalies in the SM gauge group should be satisfied as well.):

\begin{itemize}
\item Baryonic Anomalies:
\begin{itemize}
\item ${\cal A}_1 \left( SU(3)^2 \bigotimes U(1)_B \right)$: In this case all the quarks contribute and one finds
${\cal A}_1^{SM} =0$.
\item ${\cal A}_2 \left( SU(2)^2 \bigotimes U(1)_B \right)$: Since in the SM there is only one 
quark doublet, $Q_L$, for each family one cannot cancel this anomaly, ${\cal A}_2^{SM} =\frac{3}{2}$.
Therefore, here one needs extra states in a non-trivial representation of $SU(2)$.  
\item ${\cal A}_3 \left( U(1)_Y^2 \bigotimes U(1)_B \right)$: In the Abelian sector one has the contributions 
of all quarks and one finds
$
{\cal A}_3^{SM} =  - \frac{3}{2}.
$
\item ${\cal A}_4 \left(U(1)_Y \bigotimes U(1)_B^2\right)$: In this case, since all SM quarks have the same baryon number, 
this anomaly is equivalent to the $U(1)_Y$ anomaly condition in the SM, i.e. ${\cal A}_4^{SM}=0$.

\item ${\cal A}_5 \left( U(1)_B \right)$: The baryon-gravity anomaly is also cancelled in the SM, ${\cal A}_5^{SM}=0$. 

\item ${\cal A}_6 \left(U(1)_B^3\right)$: In this case one finds that this anomaly is zero in the SM, ${\cal A}_6^{SM}=0$.
\end{itemize}
\item Leptonic Anomalies:
\begin{itemize}
\item ${\cal A}_7 \left( SU(2)^2 \bigotimes U(1)_L\right)$:  It is easy to show that ${\cal A}_7^{SM}=3/2$.
\item ${\cal A}_8 \left( U(1)_Y^2 \bigotimes U(1)_L \right)$:  As in the previous case, in the SM this anomaly is not zero, 
${\cal A}_8^{SM}=-3/2$.
\item ${\cal A}_{9} \left(U(1)_Y \bigotimes U(1)_L^2\right)$: In the SM this anomaly is cancelled. 
\item ${\cal A}_{10} \left( U(1)_L\right)$: The lepton-gravity anomaly is cancelled since we add 
three families of right-handed neutrinos, $\nu_R \sim (1,1,0,0,1)$.
\item ${\cal A}_{11} \left(U(1)_L^3\right)$: As in the previous case this anomaly is cancelled once 
we add three families of right-handed neutrinos.
\end{itemize}
\end{itemize}
We have to cancel all the anomalies discussed above (and not induce standard model gauge anomalies as well) in order 
to find an anomaly free theory with B and L gauged. For a previous study of these anomalies see Ref.~\cite{Foot}.

There are two simple ways to cancel the anomalies. They are:
\begin{itemize}
\item Case 1) All baryonic anomalies are cancelled adding new quarks, $Q_L^{'T} = ( u^{'}, d^{'})_L \sim (3,2,1/6,-1,0)$, 
$u_R^{'} \sim (3,1,2/3,-1,0) $, and $d_R^{'} \sim (3,1,-1/3,-1,0)$, which transform as the 
SM quarks but with baryon number, $B=-1$. At the same time the leptonic anomalies 
are cancelled if one adds new leptons $l_L^{'T} = ( \nu^{'}, e^{'})_L \sim (1,2,-1/2,0,-3)$, 
$e_R^{'} \sim (1,1,-1,0,-3)$ and $\nu_R^{'}\sim (1,1,0,0,-3)$. All anomalies in the SM gauge 
group are cancelled since we have introduced one new full family. It differs from the usual 
standard model families since the new quarks have baryon number minus one and the new 
leptons have lepton number minus three{\footnote{For the case of baryonic anomalies see Ref.~\cite{Murayama} 
which proposes the same baryon number charge assignment for the exotic fourth generation.}}.

\item Case 2) The baryonic anomalies are cancelled adding new quarks, $Q_R^{'T} = ( u^{'}, d^{'})_R \sim (3,2,1/6,1,0)$, 
$u_L^{'} \sim (3,1,2/3,1,0) $, and $d_L^{'} \sim (3,1,-1/3,1,0)$, which transform as the 
SM quarks of opposite chirality  but with baryon number, $B=1$. At the same time the leptonic anomalies 
are cancelled if one adds new leptons $l_R^{'T} = ( \nu^{'}, e^{'})_R \sim (1,2,-1/2,0,3)$, 
$e_L^{'} \sim (1,1,-1,0,3)$ and $\nu_L^{'}\sim (1,1,0,0,3)$. All anomalies in the SM gauge 
group are cancelled since we have introduced one new full family but with opposite chirality. It also differs from the usual 
standard model families since the new quarks have baryon number one and the new 
leptons have lepton number  three. 
\end{itemize}
These are the two simplest fermionic contents that give 
an anomaly free theory with gauge group  $SU(3) \bigotimes SU(2) \bigotimes U(1)_Y \bigotimes U(1)_B \bigotimes U(1)_L$.
For the study of the phenomenological properties of models with an extra generation see Ref.~\cite{Sher}.
\section{III. Mass Generation and  Flavour Violation }
\subsection{A. Quark Sector}
We begin by considering the case 1), new generation where the chirality is the same 
as in the standard model generations. Masses for the new quarks present in the model 
are generated from their couplings to the SM Higgs:
\begin{eqnarray}
-\Delta {\cal L}_{q' {\rm mass}}^{(1)}&=& h_U^{'} \  \overline{Q^{'}_L} \  \tilde{H} \ u_R^{'} \  \nonumber \\
&+&\  h_D^{'} \  \overline{Q^{'}_L} \  {H} \ d_R^{'}   \ + \  \rm{h.c.},
\end{eqnarray}
where $H \sim (1,2,1/2,0,0)$ is the SM Higgs, and $\tilde{H} = i \sigma_2 H^*$. 

To avoid having a stable colored particle we couple the SM fermions to the new fourth generation quarks. 
However, at the same time it is important to avoid tree level flavor changing neutral currents  in the quark sector. 
We  achieve this goal by adding a new scalar field, $\phi \sim (1,2,1/2,4/3,0)$ which does not get a vev. The
interactions of this scalar are,
$
Y_1 \  \overline{Q^{'}_L} \  \tilde{\phi} \ u_R \ + \   \rm{h.c.},
$
which permit the new fourth generation quarks to decay to a scalar and a standard model quark.  
One might think that this particle is the dark matter. Unfortunately, it is not consistent with 
experiment~\cite{Drees} to have the $\phi^0$  be the dark matter because the mass degeneracy 
of the imaginary and real parts~\cite{Cirelli}. 

Next we consider case 2), where the new fermions  have the opposite chirality to the standard model. We shall see that in this case there is a dark matter candidate.  Now, mass for the new quarks is generated through the terms,
\begin{eqnarray}
-\Delta {\cal L}_{q' {\rm mass}}^{(2)}&=&Y_U^{'} \  \overline{Q^{'}_R} \  \tilde{H} \ u_L^{'} \  \nonumber \\
&+&\  Y_D^{'} \  \overline{Q^{'}_R} \  {H} \ d_L^{'}   \ + \  \rm{h.c.}.
\end{eqnarray}
Decays of the new quarks are induced by adding a new scalar field $X$ with gauge 
quantum numbers, $X \sim(1,1,0,-2/3,0)$ and the following terms occur in the Lagrange density:
\begin{eqnarray}
-\Delta{\cal L}_{{DM}}^{(2)}&=& \lambda_Q \ X \  \overline{Q_L }\ Q_R^{'} \ + \  \lambda_U \ X \  \overline{u_R} \ u_L^{'}  \nonumber \\
& + &  \lambda_D \ X \  \overline{d_R }\ d_L^{'} \ + \   \rm{h.c.}.
\label{DM}
\end{eqnarray}
The field $X$ does not get a vev and so  there is no mass mixing between the new exotic generation quarks 
and the standard model ones.  When $X$ is the lightest new particle with baryon number, it is stable.  
This occurs because the model has a global U(1)  symmetry where the $Q'_R$, $u'_L$, $d'_L$ 
and $X$ get multiplied by a phase.  This $U(1)$ symmetry is an automatic consequence of the gauge 
symmetry and the particle content.  Notice that the new fermions  have  $V+A$ interactions with the W-bosons.
The $X$ particle is a dark matter candidate and its  properties  will be investigated in a future publication. 

The field $X$ has flavor changing couplings that cause transitions between quarks with 
baryon number 1 and the usual quarks with baryon number 1/3. However, since there is no mass 
mixing between these two types of quarks integrating out the $X$ does not generate any 
tree level flavor changing neutral currents for the ordinary quarks. Those first occur at the one loop level.
\subsection{B. Leptonic Sector}
The  interactions that generate masses for the new charged leptons in case 1) are:
\begin{eqnarray}
-\Delta{\cal L}_{l}^{(1)}&=&  Y_E^{'} \  \overline{l^{'}_L} \  {H} \ e_R^{'}  \ + \  \rm{h.c.}
\end{eqnarray}
while for the neutrinos they are
\begin{eqnarray}
-\Delta{\cal L}_{\nu}^{(1)}&=&  Y_\nu \  l  H \nu^C  \ + \  Y_\nu^{'} \  l^{'} H N \ + \
\nonumber
\\
& + &  \   \frac{\lambda_a}{2} \  \nu^C \ S_L \  \nu^C \ + \  {\lambda_b} \  \nu^C \ S_L^\dagger \ N \ + \  \rm{h.c.},
\end{eqnarray}
where $S_L \sim (1,1,0,0,2)$ is the Higgs that breaks $U(1)_L$, generating masses for the right-handed neutrinos 
and the quark-phobic $Z^{'}_L$. We introduce the notation $\nu^C = (\nu_R)^C$ and $N= (\nu_R^{\prime})^C $.
After symmetry breaking the mass matrix for neutrinos in the left handed basis, $(\nu, \nu^{'}, N, \nu^C)$, 
is given by the eight by eight matrix
\begin{equation}
{\cal M}_{N} =
	\begin{pmatrix}
		0 
		&
		 0
		&
		0
		&
		M_D
	\\
		0
		&
		0
		&
		M_D^{'}
		&
		0	
	\\
		0
		&
		 (M_D^{'})^T
		&
		0
		&
		M_b
		\\
		M^T_D
		&
		0
		&
		M_b^T
		&
		M_a
	\end{pmatrix}.
\label{neutralino}
\end{equation}
Here, $M_D=Y_\nu v_H/\sqrt{2}$ and $M_a=\lambda_a v_L/\sqrt{2}$ are $3\times3$ matrices, 
$M_b=\lambda_b v_L^*/ \sqrt{2}$ is a $1\times 3$ matrix, $M_D^{'}=Y_\nu^{'} v_H/\sqrt{2}$ is a number 
and  $\langle S_L\rangle= v_L/{\sqrt{2}}$. Lets assume that the three right-handed neutrinos $\nu^C$ 
are the heaviest. Then, integrating them out  generates the following mass matrix for the three light-neutrinos :
\begin{equation}
{\cal M}_\nu = M_D \ M_a^{-1} \ M_D^T.
\end{equation}
In addition, a Majorana mass $M'$ for the fourth generation right handed neutrino $N$,
\begin{equation} 
M^{'}=M_b M_a^{-1} M_b^T,
\end{equation}
is generated. Furthermore, suppose that  $M^{'} << M_D^{'}$,  then the new fourth generation neutrinos $\nu^{'}$ and $N$ are quasi-Dirac with a mass equal to $M_D^{'}$. Of course we need this mass to be greater than $M_Z/2$ to be consistent with the measured $Z$-boson width. In this model we have a consistent mechanism for neutrino masses which is a particular combination of  Type I seesaws~\cite{TypeI}. 

The  interactions that generate masses for the new leptons in case 2) are:
\begin{eqnarray}
-\Delta{\cal L}_{l}^{(2)}&=&  Y_E^{''} \  \overline{l^{'}_R} \  {H} \ e_L^{'}  \ + \  \rm{h.c.}
\\
-\Delta{\cal L}_{\nu}^{(2)}&=&  Y_\nu \  l  H \nu^C  \ + \  Y_\nu^{''} \  \bar{l}_R^{'} \tilde{H} \nu_L^{'} \ + \
\nonumber
\\
& + &  \   \frac{\lambda_a}{2} \  \nu^C \ S_L \  \nu^C \ + \  {\lambda_c} \  \nu^C \ S_L^\dagger \ \nu_L^{'} 
\nonumber 
\\
& + &  \lambda_l \ \bar{l}_R^{'} \ l_L S_L  \ + \ \lambda_e \ \bar{e}_R \ e_L^{'} \ S_L^{\dagger}\ + \  \rm{h.c.},
\end{eqnarray}
Notice that in this case $S_L$ does not get a vev in order to avoid tree level lepton flavour violation.
Then, the neutrinos can be Dirac fermions and one has to introduce a new scalar field to break $U(1)_L$. 
Let us say $S_L^{'} \sim (1,1,0,0,n_L)$, where $n_L \neq \pm 2,\pm 6$. Notice that if in this case we do not introduce $S_L$, 
the heavy extra Dirac neutrino is stable and it is difficult to satisfy the experimental bounds from dark matter direct 
detection in combination with the collider bounds on a heavy stable Dirac neutrino. 

In order to complete the discussion of symmetry breaking we introduce a new Higgs, $S_B$, with non-zero baryon 
number (but no other gauge quantum numbers) which gets the vev, $v_B$, breaking $U(1)_B$ and giving mass 
to the leptophobic $Z^{'}_B$.  In summary, the Higgs sector in case 2) is composed of the SM Higgs, 
$S_L$, $S_L^{'}$, $S_B$ and $X$. This is the minimal Higgs sector needed to have a realistic theory where 
B and L are both gauged, and have a DM candidate. 
\section{IV. Flavour Violation and Signals at the LHC}
\begin{itemize}

\item Flavour Violation: 

Even though there are no quark flavor changing neutral currents at tree level, they do occur 
at the one loop level. The scalar $X$ is used to couple the fourth generation quarks to 
the ordinary ones and we find that at one loop there are box diagrams 
that give  contributions (after integrating out the heavy particles) to the effective Lagrangian 
for $K-{\bar K}$ mixing of the form,
  $\lambda^4 {{\bar d_{L,R}}\gamma^{\mu}s_{L,R} {\bar d_{L,R}}\gamma_{\mu}s_{L,R} /(16 \pi^2 M^2}) +{\rm h.c.}$,
where the product of the four elements of the Yukawa matrix $\lambda_i$ (and CKM angles) that enter into the coefficient are 
denoted by $\lambda^4$ and $M$ denotes a mass scale set by the masses of the new fields in the loop. For $M$ of order 
$100 ~{\rm GeV}$ this is negligible  provided $\lambda <10^{-2}$. 

Charged lepton neutral currents are induced at one loop level. For example, there is a one loop contribution to the 
amplitude for $\mu \rightarrow e \gamma$. It involves the usual factor of the muon mass and one loop suppression 
factor.  In addition it requires two factors of the  mixing between the essentially massless ordinary neutrinos and 
the new fourth generation neutrino. In the limit we discussed above where the fourth generation neutrino in 
quasi-Dirac this mixing is small.  A detailed study of the flavour violation issue in this context is beyond the scope of this letter.

\item $Z_L$ and $Z_B$:

In this model one can observe lepton number violating processes 
at the LHC through the channels with same-sign dileptons:
$ pp \  \to \ Z_L^* \  \to \  \nu_R \ \nu_R \  \to \  W^{\pm} \  W^{\pm} \  e^{\mp}_i  \  e^{\mp}_j$.
However, for this Drell-Yan production one needs the mixing between 
$Z$ and $Z_L$, or $Z_L$ and $Z_B$. One can also have pair production of  
$Z_L$ through its couplings to the physical Higgses.
For a study of these channels see Ref.~\cite{Tao}. 
In the case of the lepto-phobic $Z_B$  its coupling to the third 
generation of quarks can be used to observe it at the LHC. 
In particular, the channel:
$pp \  \to \  Z_B^* \  \to \  t  \  \bar{t}$.
For previous studies of a lepto-phobic $Z^{'}$ see Ref.~\cite{ZB}.
For a review on the properties of  new $U(1)$ gauge bosons 
see for example Ref.~\cite{Langacker}.

\item Decays of the new quarks:

As we have discussed above  to avoid a stable colored 
particle we introduced the new interactions in Eq.(\ref{DM}).
Now we discuss signals at the LHC coming from these 
interactions. Here we will focus on the case where the 
new quarks can decay into $X$ and the top quark. 
Then, one can have the interesting  channel:
$pp \  \to \  \bar{t}^{'}  \  t^{'} \  \to \  X \  X \  \bar{t} \ t$.
Since $X$ is stable, the final state has missing energy 
and a $t \bar{t}$ pair (After posting the first version of 
this article the paper~\cite{Feng} appeared.). 

\item Baryon number violating processes:

Since baryon number is spontaneously broken it is important to 
consider possible baryon number violating operators that might 
arise after symmetry breaking. Using the minimal Higgs sector 
discussed above one can see that the operator, $QQQL/\Lambda_B^2$, 
is never generated since we do not have a Higgs which carries B and L 
quantum numbers.  Also, in general we expect the $\Delta B=2$ operators 
only when $S_B$ has $B=2$, but in general we do not induce this operator as well.
Hence the model we have introduced does not give rise to dangerous 
baryon number operators. At the same time, in order to avoid the vev 
for $X$, one should forbid the terms with an odd power of $X$ in the scalar potential. 
Then, one must impose the condition $B(S_B) \neq \pm 2/3, \pm 1/3, \pm 2/9, \pm 2$.

\item Other aspects:

It is well known, that in any model with an extra fermionic generation 
one finds that the gluon fusion cross section for the SM Higgs is larger 
by a factor 9. See for example Ref.~\cite{Tilman}. However, the new results from CDF and D0~\cite{Higgs}, 
do not rule out our model when the Higgs mass is $114 \  \rm{GeV} \  <  \  M_H \  < 120$ 
GeV, or when $M_H > 200$ GeV.  Notice that for a large mixing between $H$ and the singlets 
$S_L$ and $S_B$ one can relax those constraints. For the study of other aspects, 
such as electroweak precision constraints, see for example Ref.~\cite{Tilman}.

\end{itemize}
\section{V. Summary}
We have constructed and investigated  a simple theory where Baryon (B) number and Lepton (L) number 
are local gauge symmetries. In this theory, B and L are treated on the same footing,  anomalies are cancelled 
by adding a single new fermionic generation, there is  a simple realization of the seesaw mechanism 
for neutrino masses and there is a natural suppression for flavour violation at tree level 
in the quark and leptonic sectors. It is important to emphasize that in this theory the B and L violation 
scales can be as low as TeV and one does not have dangerous processes, such as proton decay. 
Also one finds that the stability of a dark matter candidate is an automatic consequence of the gauge symmetry.
Constraints and signals at the Large Hadron Collider have been briefly discussed.  
The results of our paper can be applied to different theories 
for physics beyond the Standard Model, such as the Minimal Supersymmetric 
Standard Model, where new interactions can give rise to fast proton decay. 

\subsection*{Acknowledgment}
P. F. P. would like to thank T. Han and S. Spinner for useful discussions and pointing out Ref.~\cite{Feng}.
The work of P. F. P. was supported in part by the U.S. Department of Energy
contract No. DE-FG02-08ER41531 and in part by the Wisconsin Alumni
Research Foundation.  The work of M.B.W.  was supported in part by the U.S. Department of Energy under contract No. DE-FG02-92ER40701. 


\end{document}